\documentclass[journal]{IEEEtran}
\usepackage[noadjust]{cite}
\usepackage[dvips]{graphicx}
\usepackage{ushort}
\usepackage{psfrag}
\usepackage{epsf}
\usepackage{bm}
\usepackage{mathtools}
\usepackage{trfsigns}
\usepackage{amsmath}
\usepackage{amssymb}

\usepackage{enumitem}
\usepackage{color}

\begin{document}
\graphicspath{{/}}

\title{Use of Correlation Matrix to Assess the Stirring Performance of a Reverberation Chamber: a Comparative Study}
\author{%
        Gabriele~Gradoni,~\IEEEmembership{Member,~IEEE},
        Valter~{Mariani~Primiani},~\IEEEmembership{Member,~IEEE},
        Franco~Moglie,~\IEEEmembership{Senior Member,~IEEE}
\thanks{Manuscript received Xxxxxxx xx, xxxx; revised Xxxxxxxx xx, xxxx.}
\thanks{
G.~Gradoni is with Institute for Research in Electronics and Applied Physics,
University of Maryland,
College Park, MD 20742 USA
(e-mail: ggradoni@umd.edu).
F.~Moglie and V.~{Mariani~Primiani} are with Dipartimento di Ingegneria dell'Informazione,
Universit\`a Politecnica delle Marche,
via Brecce Bianche 12, 60131 Ancona,
Italy (e-mail: f.moglie@univpm.it; v.mariani@univpm.it).}}
\markboth{IEEE Transactions on Electromagnetic Compatibility,~Vol.~XX, No.~X, XXX~2014}%
{Gradoni \MakeLowercase{\textit{et al.}}: XXXXXXXXX}
\maketitle
\begin{abstract}
The use of correlation matrices to evaluate the number of uncorrelated stirrer positions 
of reverberation chamber has widespread applications in electromagnetic compatibility. 
We present a comparative study of recent techniques based on multivariate correlation 
functions aimed at relating space-frequency inhomogeneities/anisotropies to the reduction of 
uncorrelated positions. Full wave finite-difference time domain simulations of an actual 
reverberation chamber are performed through an in-house parallel code. 
The efficiency of this code enables for capturing extensive inhomogeneous/anisotropic 
spatial volumes (frequency ranges). 
The concept of threshold crossing is revised under the light of random field sampling, 
which is important to the performance of arbitrary reverberation chambers. 
\end{abstract}


\section{Introduction}
\label{sec:introduction}

Performances of a mode-stirred reverberation chamber (RC) in electromagnetic compatibility (EMC) 
applications are intimately related to the number of independent cavity realizations \cite{iec-61000-4-21}.
Unambiguous evaluation of the number of independent stirrer positions 
is still under investigation \cite{2012_EMC_EUROPE_Pfennig}.
Current approaches rely  
on the autocorrelation of fields sampled at single site inside the working volume (WV) of the RC. 
The coherence time is based on the autocorrelation function \cite{2009_Sorrentino_AP_Coherence}.
Previous investigators showed that, despite inside the WV,
by considering only one chamber site results in a high spatial variability of the 
autocorrelation coefficient \cite{besnier_corr_2011, RC_2012_Chen, RC_2013_Pirkl}.
This experimental evidence is of crucial importance as the commonly accepted notion of independence
for a stirrer position is just defined by the 
$\rho_e = e^{-1}$ threshold crossing of the autocorrelation coefficient. 
Even though it brings about a simple and effective criterion,
this perspective is provably incomplete and uncertain in evaluating 
whether members of the cavity ensemble is strictly ``independent'' each other. 
To this regard, the threshold $\rho_e$ is typically associated with the concept of ``uncorrelation'' rather than ``independence''.  
Those inspections lead to consider a correlation matrix
having each entry defined by the correlation between two of the total $N_s$ stirrer positions. 
Of course, the dimension of the so defined correlation matrix is given by $N_s \times N_s$.
In a good RC, we expect a low value for many correlation matrix elements,
i.e. many stirrer position pairs are ``low correlated''.

According to this multivariate approach, an alternative way of evaluating
the number of uncorrelated positions of mechanical mode-stirrers
 is based on the calculation of the correlation matrix through 
a grid of $N_w$ spatial points,
selected among an arbitrary volume of the chamber \cite{2013gradoni_pier_correl}.
Another recent application of the correlation matrix
involves single-point measurements through a wide frequency and to evaluate ``uncorrelated frequencies'' 
rather than spatial points \cite{RC_2012_pirkl}.
This way proves more fast and efficient as populating spatial correlation matrix requires for 
a large amount of field/power measurements. 
Hybrid techniques can also be conceived based on the correlation matrix. 
Spatial point and frequency point data can be combined
to check stirring performance. 

In this work we review existing multivariate strategies grounded on the correlation matrix that are important to assess RC performances.
The typical framework we use to test different performance indicators is a reverberation chamber equipped with a ``carousel'' 
mode-stirrer \cite{RC_2012_Moglie}.
Being based on multispatial multifrequency sampling, these indicators can be used in arbitrary dynamic and complex wave environment. 
In particular, the mode stirrer can be either present as in RCs or absent as in wave chaotic billiards or enclosures \cite{1998_Stoeckmann_Sinai3D, 2005_Legrand_S}. 
Observations are useful to explore the physics of electromagnetic reverberation beyond overmoded regime, and to create finite sets of 
realizations behaving closely to ideal statistical ensembles. There is room to think that judicious multi-resolved field sampling could 
lead to a lower chamber lowest usable frequency (LUF) \cite{iec-61000-4-21}. 

The strategies being used through the rest of the paper are focused on the evaluation of:
\begin{enumerate}
\item uncorrelated stirrer positions adopting a spatial correlation matrix (henceforth referred as ``US--PM'');
\item uncorrelated stirrer positions adopting a frequency correlation matrix (henceforth referred as ``US--FM'');
\item uncorrelated frequency steps adopting a spatial correlation matrix
      for a single stirrer position (henceforth referred as ``UF--PM'').
\end{enumerate}

In any strategy we analyze the effect of varying the involved parameters,
number and distance of frequency and spatial points,
and we compare results to those obtained applying the 1--D circular autocorrelation method. 

Finally, of interest in RC theory, we propose and use strategies for the analysis of:
\begin{enumerate}[resume]
\item the uncorrelated frequency steps adopting a stirrer position correlation matrix
      at a single spatial point (henceforth referred as ``UF--SM'');
\item uncorrelated spatial points adopting a stirrer position correlation matrix
      for each single frequency (henceforth referred as ``UP--SM'');
\item uncorrelated spatial points adopting a frequency correlation matrix (henceforth referred as ``UP--FM'').
\end{enumerate}

 


\begin{figure}[!t]
\psfrag{x}{\scriptsize{x}}
\psfrag{y}{\scriptsize{y}}
\psfrag{z}{\scriptsize{z}}
\psfrag{Lx}{\scriptsize{$L_x$}}
\psfrag{Ly}{\scriptsize{$L_y$}}
\psfrag{Lz}{\scriptsize{$L_z$}}
\psfrag{Hs}{\scriptsize{$h$}}
\psfrag{De}{\scriptsize{$d_e$}}
\psfrag{Di}{\scriptsize{$d_i$}}
\centering
\includegraphics[width=2.5in]{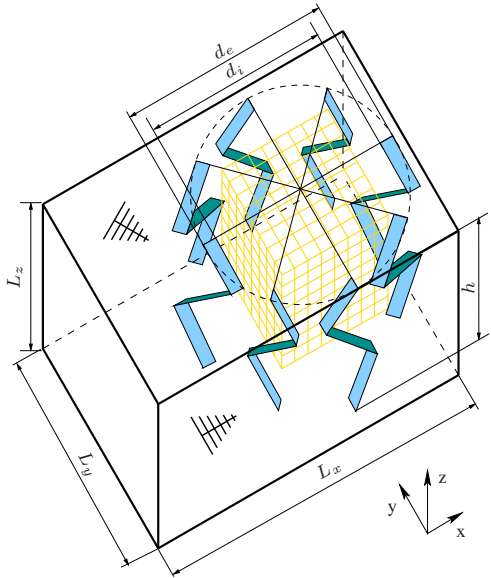}
\caption{Geometry of the reverberation chamber (RC) equipped with a ``carousel'' stirring system.
         The simulated RC is that of Ancona EMC laboratory, it is $L_x=6$~m long, $L_y=4$~m wide and $L_z=2.5$~m height. 
         The vertical plates are $h=2.4$~m high and they have a Z-folded shape. 
         The rotating system describes a cylindrical volume whose base is centered in $x = 4$~m and $y = 2$~m, the external diameter is $d_e=3.8$~m and the internal diameter is $d_i=3.3$~m.}
\label{fig:geocamera}
\end{figure}

\section{RC Performance Indicators}
\label{sec:indicators}
The use of the circular autocorrelation function (ACF) is well accepted for evaluating the number of independent 
stirrer positions \cite{iec-61000-4-21}. More advanced methods based on the ACF have been proposed to assess the 
performances of mode-tuned reverberation chambers (MTRC) \cite{besnier_goftest_2007, besnier_corr_2011}.
Recently, the concept of multi-resolution for the construction of correlation matrices in MTRC is attracting the interest 
of the scientific community \cite{2011_Cozza_York}

The definition of independent stirrer positions based on a multivariate field perspective has been pioneered in
\cite{2012_EMC_EUROPE_Pfennig, 2013gradoni_pier_correl}.
In \cite{RC_2012_Chen}, the number of uncorrelated positions are related to the number of independent eigenvalues 
of the correlation matrix $\ushortdw{R}_s$. 
In particular, $\ushortdw{R}_s$ is populated by Pearson correlation coefficients (PCC). 
Those elements are calculated as two-point correlation functions between two arbitrary points, and 
therefore require for multi-resolved field sampling. 
This has been achieved through a ``platform stirring'' strategy \cite{RC_2012_Chen}. 
From information theoretic arguments, it turns out that the number of independent eigenvalues is 
\begin{equation}\label{eqn:N_traceR}
 N_{\rm eff} = \frac{\textrm{Tr}^2 \left [ \ushortdw{R}_s \right ]}{\textrm{Tr} \left [ \ushortdw{R}_s^2 \right ]} \,\, .
\end{equation}
This perspective has the advantage to be threshold-less, i.e.,
the elements of $\ushortdw{R}_s$ do not need to be tested against the empirical $e^{-1}$ limit to 
achieve $N_{\rm eff}$. 
Threshold-based approaches currently adopted in the IEC normative \cite{iec-61000-4-21}
rely on single point circular autocorrelation coefficient (CC) to evaluate $N_{\rm eff}$.
With respect to the IEC method, usage of (\ref{eqn:N_traceR}) underestimates $N_{\rm eff}$.

A similar philosophy is adopted in \cite{RC_2012_pirkl} where the concern is the number of uncorrelated measurements
rather than stirrer positions. 
The measurement correlation has been defined in terms of the maximum Reny\`i entropy, again based on the eigenvalue of the correlation 
matrix. 
In particular, given the maximum entropy achievable through $N_s$ measurements
\begin{equation}\label{eqn:maxentr}
 S_2 = \log \left ( N_{\rm eff} \right ) \,\, ,
\end{equation}
it is straightforward to have an estimation of the 
number of uncorrelated measurements reading 
\begin{equation}\label{eqn:Mhat}
 N_{\rm eff} = e^{S_2} \,\, .
\end{equation}
Since $S_2$ can be calculated from the eigenvalues of $\ushortdw{R}_s$, then 
\begin{equation}\label{eqn:meff_eigen}
 N_{\rm eff} = \frac{\left ( \sum_{m=1}^{N_s} \lambda_m \right )^2}{\sum_{m=1}^{N_s} \lambda_m^2} \,\, .
\end{equation}
The analogies between between (\ref{eqn:meff_eigen}) and (\ref{eqn:N_traceR}) are many,
and essentially they constitute the same formula achieved through different perspectives \cite{1990_Widmann_DoF}. 
The advantage of (\ref{eqn:meff_eigen}) is that the correlation matrix can be populated through a frequency scanning,
i.e., a frequency sampling would lead to $\ushortdw{R}_f$ as in \cite{RC_2012_pirkl},
rather than the spatial scanning, leading to $\ushortdw{R}_s$.
It is worth remarking that the estimator (\ref{eqn:N_traceR}), used in different versions in \cite{RC_2012_pirkl} and \cite{RC_2012_Chen},
is polarized, and the estimated number of uncorrelated positions $N_{\rm eff}$, based on $M$ space/frequency points, is given by 
\begin{equation}\label{eqn:N_bias}
 \frac{N_{\rm eff}}{N_s} = \frac{1}{1 + \frac{N_s}{M}} \,\, ,
\end{equation}
which converges to $N_{\rm eff}$ for a relatively high number of degrees of freedom $M$, i.e., $M \gg N_s$. 
Equation (\ref{eqn:N_bias}) can be demonstrated by noting that the correlation matrix
can be expanded in true principal components (PC) \cite{1990_Widmann_DoF}. 
Inherently, it has been recently demonstrated that the RC field admits PC decomposition
of the stirring process in empirical (time-domain) modes, called as 
``eigenstirrings'' \cite{RC_2013_Arnaut2}.
Therefore, (\ref{eqn:N_bias}) can be also used to perform a systematic correction of the estimates if we make use 
of (\ref{eqn:N_traceR}) to evaluate the number of effective stirrer positions
with a few finite difference time domain (FDTD) lattice points \cite{2013gradoni_pier_correl}.

Based on an entropic perspective, we can use (\ref{eqn:meff_eigen}) to carry out the calculation of uncorrelated stirrer positions 
through either $\ushortdw{R}_s$ or $\ushortdw{R}_f$. It has been already acknowledged that the production of entropy 
from scattering processes over finite regions of space is important to evaluate the degrees of freedom of a complex EM system
\cite{1989_Bucci_freedom}.
A thorough study of the time-domain entropy generation in RC has been performed in \cite{2012_EMCEurope_Entropia},
through numerical FDTD simulations of an actual RC,
and the linear increase of entropy at early times was proven to saturate at the maxentr limit (\ref{eqn:maxentr})
which converges to the high-frequency estimate.
This behavior is typical of multi-component (many-body) complex systems and has been observed in quantum many-body systems operating in 
chaotic regime \cite{2012_Borgonovi_PhysRevLett}.

A step forward has been carried out in \cite{2013_pfennig_brugge} through the evaluation of uncorrelated pairs in the correlation matrix.
This method grasps on graph theory to 
estimate the number of uncorrelated stirrer positions. 
A graph is developed where each node corresponds to a stirrer position,
and an arc is formed when the correlation between two positions $(i,j)$ 
satisfies the cutoff condition \cite{iec-61000-4-21}
\begin{equation}\label{eqn:impr_limit}
 r_{ij} = \left [ \ushortdw{R}_s \right ]_{ij} \leq \e^{-1} \left [ 1 - \frac{7.22}{\left ( N_s^2 \right )^{0.64}} \right ] \,\, .
\end{equation}
Equation (\ref{eqn:impr_limit}) is an extension of the correlation threshold $\rho_e$ 
based on sampling theory. 
The author then uses standard algorithms to search for the so-called ``maximum clique'' of the graph,
and identifies it as the maximum number of uncorrelated positions.
The result is a graphical method to count the number of uncorrelated positions through a direct test of uncorrelated node pairs. 
The method in \cite{2012_EMC_EUROPE_Pfennig, besnier_corr_2011} relies on a threshold,
while the methods proposed in \cite{RC_2012_Chen, RC_2013_Pirkl} do not rely on it, i.e., they can be classified as threshold-less. 

The presence of a threshold should not be taken necessarily as a limitation.
Actually, its physical significance will become clear in a random sampling perspective, where 
we imagine to populate $\ushortdw{R}_{s,f}$ with fluctuating elements. 
Following this philosophy, we decided to introduce a generalized threshold-based procedure
to estimate the number of uncorrelated positions, by exploiting 
information of an extended reverberation (sub) space rather than a single point.
The generality of the method resides in the fact that the sampling over multiple spatial points/frequencies has been extended to
an arbitrary reverberant subspace independently on the a-priori definition of a working volume, and beyond the orientation of the DUT. 
The mathematical procedure is detailed in \cite{2013gradoni_pier_correl}.
The number of uncorrelated stirrer positions comes from counting the number of elements in $\ushortdw{R}_{s,f}$ that are below the 
threshold (\ref{eqn:impr_limit}). Each element has the meaning of a correlation between two cavity sub-spaces (calculated through a 
discrete space/frequency lattice \cite{2013_EMCS_Correlazione} generated from two different stirrer configurations. 
More precisely, this number is obtained by taking account of the symmetry properties
of $\ushortdw{R}_{s,f}$, and of the fact that $r_{ii} = 1$, hence
\begin{equation}\label{eqn:Nind}
 N_{\rm eff} = \frac{N_s^2}{\# \left [ \ushortdw{R}_{s,f} > r  \ushortdw{1} \right ]} \,\, ,
\end{equation}
where 
$\# \left [ \cdot \right ]$ is the counting operator,
$\ushortdw{1}$ is the a square matrix of dimension $N_s$, where all elements are 1.
When two stirrers are used in an RC, we follow the procedure described in \cite{Moglie_TEMC_2011}.
Here, the case involving $\ushortdw{R}_s$ corresponds to US--PM, while the case involving 
$\ushortdw{R}_f$ corresponds to US--FM. 

Interestingly, besides stirrer positions, our method can be used
to evaluate the number of uncorrelated spatial points  given measurements from a 
set of stirrer positions (US--SM, UP--FM).
Also, the number of uncorrelated working frequencies can be estimated from stir sequences 
(UF--SM), or spatial lattice sampling (UF--SM).

\section{Full-wave simulations and results}
\label{sec:simres}
We now evaluate the number of uncorrelated stirrer positions of an actual RC. 
In particular, we focus on the stirring performances of the ``carousel'' stirrer 
previously studied in \cite{RC_2012_Moglie}.
\figurename~\ref{fig:geocamera} shows the detailed geometry: the stirrer consists of equispaced metallic $z$-folded blades. 
Its rotation describes a cylindrical 
volume which bounds the uniformity volume of the chamber, i.e., the working volume (WV).

According to the method discussed in Section~\ref{sec:indicators},
we do not restrict our investigations to an \emph{a-priori} defined 
working volume, further requiring a calibration protocol such as that describe in \cite[A.5]{iec-61000-4-21}.

Full wave FDTD simulations of the RC have been recently used by other investigators
\cite{picon_FDTD_2008, RC_2013_adardour, RC_2014_Cui}.
An in-house FDTD code, optimized for BlueGene computer architecture \cite{RC_2013_mariani, RC_2013_Moglie},
is used to perform a full wave simulation of the Ancona's RC. 
The EM fields can then be sampled over a dense grid of spatial points (sampling lattice),
for an arbitrary number of stirrer positions \cite{2013gradoni_pier_correl}.

The RC is then discretized in $201 \times 134 \times 84$ cubic cells with a side of $30$ mm,
the FDTD time step is $\Delta t =50$~ps, and the number of 
iterations necessary to analyze the entire structure is found to be $206\,748$.
Frequency data are obtained by fast Fourier transform (FFT).
\figurename~\ref{fig:geocamera} shows the sampling lattice (gold solid line grid) adopted in the simulations,
where the $N_{w}$ points are 0.15~m equispaced.

The discrete (total or Cartesian) field $E_{(m_i,m_j,m_k)}^{\left ( \tau \right )}$ is picked up
at $N_w$ spatial points 
$(m_i \Delta x, m_j \Delta y, m_k \Delta z)$ of the sampling grid,
for each stir state $\tau_i = i \Delta \theta$, $i = 1, \ldots, N_s$,
with $\Delta \theta$ angular stirrer step, and $N_s$ total number
of stirrer positions considered in the analysis. 
In discretizing the continuous stirring, we assumed that the ``uncorrelation angle''
of the carousel stirrer is greater than $\Delta \theta$.
The correlation is then computed by using $N_p$ field values.
In particular, for the total electric field we have $N_p=N_w$,
while for the three separate Cartesian components we have $N_p=3N_w$.

The RC is simulated for $512$ (equispaced) stirrer angles,
and for each angle the three Cartesian field components are computed in a grid of $8 \times 8 \times 8 = 512$.
All the above field data were saved for $2622$ frequency points in the range $0.2$--$1.0$~GHz. 
The memory required to store all the data is $189$ GBytes. 


\subsection{Quality factor and coherence bandwidth}
\label{sec:qfactor}
\begin{figure}[!t]
\centering
\includegraphics[width=2.5in]{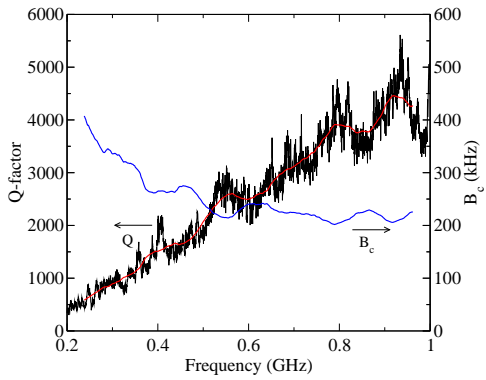}
\caption{Simulated Q-factor and coherence bandwidth.}
\label{fig:fattq}
\end{figure}

We start our analysis by computing the quality factor of the Ancona's RC in the investigated frequency range. 
An average quality factor (Q) can be derived directly from simulated scattering data by using the formula \cite{1994_hill_Q}
\begin{equation}
 Q = \frac{16 \pi^2 V \left < \left | S_{21}\right |^2 \right >}{\eta_{Tx} \eta_{Rx} \lambda^3 \left ( 1 - \left | \left < S_{11} \right > \right |^2 \right )} \,\, ,
\end{equation}
where $V$ is the RC volume, $\lambda$ the free-space wavelength, $S_{21}$ the complex scattering transmission,
and $S_{11}$ the complex scattering reflection coefficient, 
$\eta_{Tx}$ and $\eta_{Rx}$ the transmitting and receiving antenna efficiency respectively.  
In case of lossless and load matched antennas, $\eta_{Tx} = \eta_{Rx} = 1$. 
An accurate estimate of $Q$ allows for recovering
the RC coherence bandwidth $B_c$ \cite{ChenAWPL_2009_5072260, 2008_delangre_el, RC_2012_Holloway} 
\begin{equation}
 B_c = \frac{f}{\left < Q \right >} \,\, .
\end{equation}
\figurename~\ref{fig:fattq} shows the simulated quality factor and the corresponding coherence bandwidth.
It is worth noticing that for working frequencies $f > 350$~MHz,
the RC coherence bandwidth becomes lower than the sampling frequency step we use, which 
is $\Delta f \approx 305$~kHz.
This regime corresponds to high-quality factor modes accompanied by a high modal density which is predicted by 
the Weyl law \cite{2001_Luk_low_frequency}.
Despite in high frequency regime, we expect for spectral rigidity and avoided level crossing to cause overlapping
between non-degenerate modes \cite{1998_Stoeckmann_Sinai3D, 2005_Legrand_S, Cozza_TEMC_2011}.

\subsection{Uncorrelated stirrer positions}
\label{sec:resuncpos}
\begin{figure}[!t]
\centering
\includegraphics[width=2.5in]{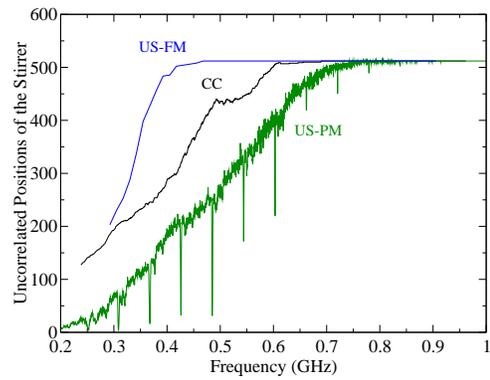}
\caption{Simulated uncorrelated stirrer positions, for a correlation matrix of grid points (US--PM) and 
         of frequency points (US--FM). Results applying the IEC standard method are also reported (CC).}
\label{fig:UncStirr}
\end{figure}

Referring to the notation of Section~\ref{sec:introduction},
we compute $N_{\rm eff}$ either through sampling grid of $N_w$ spatial points, i.e., the case 1) US--PM, and through a frequency 
scanning over $N_f$ frequency points, i.e., the case 2) US--FM. 
\figurename~\ref{fig:UncStirr} shows $N_{\rm eff}$ as calculated with $N_w=512$ and $N_f=512$.
The same Figure also reports $N_{\rm eff}$ computed through the standard, i.e., circular correlation (CC), 
technique as described in the IEC standard \cite{iec-61000-4-21}.
It is worth noticing that the US--PM method underestimates $N_{\rm eff}$ with respect to the CC method.
This happens also varying the number of grid spatial points \cite{2013_EMCS_Correlazione}
and the number of frequency points \cite{2013_EMCEurope_Correlazione}.
Conversely, the US--FM method overestimates $N_{\rm eff}$. 
Again, the sampling frequency step is greater than or equal to $B_c$.
The effect of bandwidth variation was investigated in \cite{2013gradoni_pier_correl},
where a lower number of frequency points were adopted to build up the correlation matrix. 

Interestingly, the curve related to US--FM saturates when the cavity coherence bandwidth reaches the sampling frequency, viz., 
\begin{equation}
 B_c \approx \Delta f = 305 \mbox{ kHz} \,\, .
\end{equation} 
Below this frequency, evaluating uncorrelated positions with this strategy mean sampling over $B_c$, 
as shown in \figurename~\ref{fig:fattq}, whence $B_c < \Delta f$ results in $N_{\rm eff} \leq N_{s}$. 

\subsection{Uncorrelated frequency points}
\label{sec:rescaso_uf}
\begin{figure}[!t]
\centering
\includegraphics[width=2.5in]{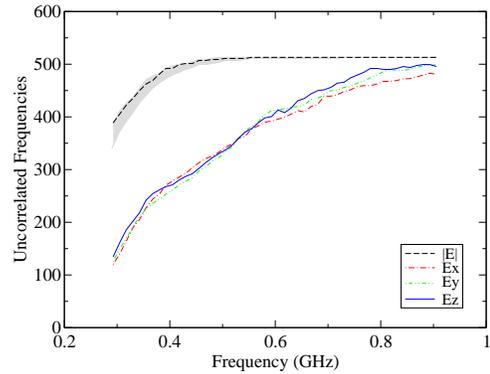}
\caption{Simulated uncorrelated frequency points for a correlation matrix of grid points (UF--PM).
         Each component and the magnitude of the electric field are reported.
         The gray area is the spread for 16 stirrer angles.}
\label{fig:ufpm}
\end{figure}

In those RC applications, where the electronic stirring method is thus employed, 
it is important to know whether the frequency steps are uncorrelated or not.
\figurename~\ref{fig:ufpm} shows the number of uncorrelated frequencies as calculated 
in the UF--PM strategy. 
It is worth pointing out two interesting features. 
Adopting the total field $|E|$, the frequency points become correlated below 400~MHz, where the frequency step is 
lower than the chamber $B_c$. 
On the other hand, the single rectangular component gives a lower number of frequency uncorrelation points.
Conversely, there is no difference when the uncorrelated stirrer positions are evaluated by $|E|$ or
by $E_{x,y,z}$ \cite{2013_EMCS_Correlazione}, i.e. in the US--PM strategy.
Results refer to a fixed stirrer angle.
The gray area in \figurename~\ref{fig:ufpm} shows the spreading of the uncorrelated frequency points 
for 16 stirrer positions. This indirectly quantifies the uncertainty of the estimation. 
\figurename~\ref{fig:ufpm} shows a very limited spreading. 

\begin{figure}[!t]
\centering
\includegraphics[width=2.5in]{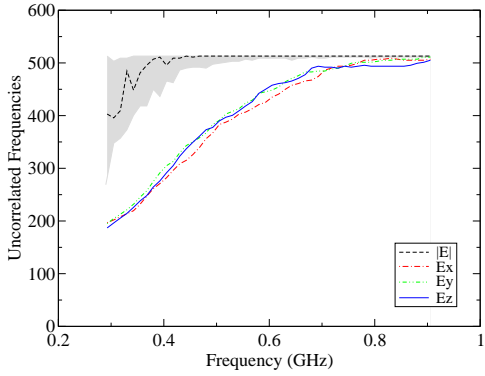}
\caption{Simulated uncorrelated frequency points for a correlation matrix of stirrer angles (UF--SM).
         The gray area is the spread for all the 512 grid points.}
\label{fig:ufsm}
\end{figure}

The determination of uncorrelated frequency number can also be done populating the correlation matrix 
with the field values computed in a single chamber point by rotating the stirrer (UF--SM).
\figurename~\ref{fig:ufsm} shows the simulated number of uncorrelated frequencies.
A similar behavior to the case UF--PM is observed.
The gray area of \figurename~\ref{fig:ufsm} shows a larger uncertainty than the one in \figurename~\ref{fig:ufpm}.

The discrepancy of \figurename~\ref{fig:ufpm} and \figurename~\ref{fig:ufsm} can be ascribed to the nature of modes inside of an RC.
The Weyl law quantifies the eigenmode density of an EM cavity, while 
no information is given on the nature of those eigenmodes.

At relatively low frequencies, the chamber modes (TE and TM) can be still considered unperturbed \cite{iec-61000-4-21}.
At higher frequencies, the modes become hybrid due to the presence of stirrer blades,
and all the three Cartesian components are involved in the total field representation.
Uncorrelated stirrer positions are not affected by any 
partitioning since their evaluation involves averaging over space (mode topology)
or over frequency (eigenspectrum) of the RC fields. 

\subsection{Random stir sampling} 
\label{sec:resrandomsampling}
The definition (\ref{eqn:Nind}) has a straightforward physical meaning:
the number of independent positions can be evaluated as the number  
of stirring configurations having acceptable spatial dispersion of the (Cartesian or total) field variance.
Therefore, given a sampling subspace, it is possible to construct a two-position correlation function 
over a discretized, i.e., tuned, stirrer rotation. 
The theoretical prediction of $N_{\rm eff}$ based on (\ref{eqn:Nind}) is not an easy task. 

In a very general fashion, we could also treat $r_ij$ as a random variable through $\ushortdw{R}_{s,f}$.
In this perspective, its probability distribution\footnote{meaning frequency of value occurrence in the Laplace sense}
would be regulated by the space/frequency sampling. 
Inherently, a randomized sampling over continuous stirrer rotations is more realistic
and would naturally call for a statistical treatment of $r_{ij}$, though in principle 
also in deterministic sampling it is possible to linearize $\ushortdw{R}_{s,f}$, order its elements $r_{ij}$, and calculate 
the probability density function (PDF) of the correlation as a distribution of its values. 
Therefore, also $N_{\rm eff}$ becomes a random variable with probability density function $f_{N_{\rm eff}} \left ( n_{eff} \right )$. 
Exploiting the addition theorem in probability,
we can turn the probability of the counting operator into a union of probabilities of the single element 
$f_{R} \left ( r < \overline{r} \right )$. 
Obviously, the ability of obtaining many single uncorrelated elements $r=r_{ij} \leq \overline{r}$
depends on the stirring parameters such as structure geometry, location, and dimensions.
The evaluation of the stirring efficiency can be performed deterministically
by full-wave numerical simulations or measurements on actual 
structures. Our perspective calls  for a statistical approach to the problem:
instead of a pure deterministic evaluation, it would be useful to introduce a probability 
$f_{R} \left ( r < \overline{r} \right )$ of the mode-stirrer to make two arbitrary realizations uncorrelated. 

Furthermore, counting $n_{eff}$ positions in $\mathcal{N}_s$ elements means having $n_{eff}$ successes
with probability $f_{R} \left ( r < \overline{r} \right )$ 
in $\mathcal{N}_s$ trials, that is given by the Bernoulli (or binomial) distribution.
Assuming \emph{uncorrelated elements} of the correlation matrix, and a deterministic threshold of $\overline{r}$, yields
\begin{equation}\label{eqn:f_ind_r}
 \begin{split}
 f_{N_{\rm eff}} \left ( n_{eff} , \overline{r} \right ) = \binom{n_{eff}}{\mathcal{N}_s} \, \left [ f_{R} \left ( r < \overline{r} \right ) \right ]^{n_{eff}} \\
  \left [ 1 - f_{R} \left ( r < \overline{r} \right ) \right ]^{\mathcal{N}_s - n_{eff}} \,\, ,
 \end{split}
\end{equation}
where the effective number of considered pairs of chamber configurations is given by half the number
of off-diagonal elements of the correlation matrix, viz., 
\begin{equation}
 \mathcal{N}_s = \frac{N_s \left ( N_s - 1\right )}{2} \,\, ,
\end{equation}
and the event probability is actually a cumulative distribution function (CDF)
\begin{equation}
 F_{R} \left ( \overline{r} \right ) =  f_{R} \left ( r < \overline{r} \right ) = \int_0^{\overline{r}} \,\, f_{R} \left ( r \right ) \,\, d r \,\, .
\end{equation}

Further physical considerations would be involved in presence of lattice autocorrelation. 
In this paper we treat the case of uncorrelated sampling space/frequency lattice.
Once obtained the distribution of $n_{eff}$, a robust estimation of the number of independent is possible by 
continuous averaging
\begin{equation}
 \overline{N}_{\rm eff} \left ( \overline{r} \right ) = \int \,\, \delta \left ( n_{eff} - n \right ) \, f_{N_{\rm eff}} \left ( n, \overline{r} \right ) \, d n \,\, , 
\end{equation} 
which is given by 
\begin{equation}
 \overline{N}_{\rm eff} \left ( \overline{r} \right ) = \mathcal{N}_s F_{R} \left ( \overline{r} \right ) \,\, .
\end{equation}
The challenge is now devoted to the derivation of the correlation distribution $f_{R} \left ( r \right )$,
and to understanding its dependence on 
chamber and stirrer parameters \cite{Moglie_TEMC_2011, RC_2012_Mariani}
such as chamber volume and losses. 
Here, we present distributions the correlation function resulting from numerical (FDTD) data. 

\begin{figure}[!t]
\centering
\includegraphics[width=3in]{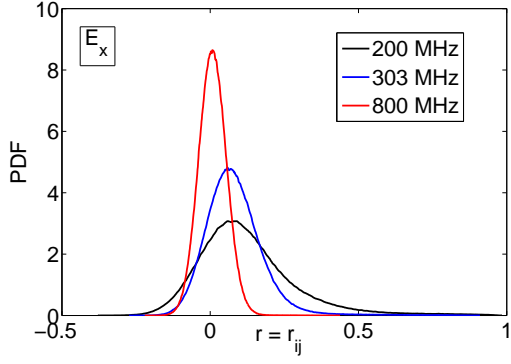}
\caption{Probability density function of the off-diagonal element of correlation matrix for $\left | E_x \right |$ at low (200~MHz),
intermediate (303~MHz), and high (800~MHz) frequency, for a spatial grid $13 \times 13 \times 11$ of the US--PM strategy.}
\label{fig:Ex_rij}
\end{figure}

\begin{figure}[!t]
\centering
\includegraphics[width=3in]{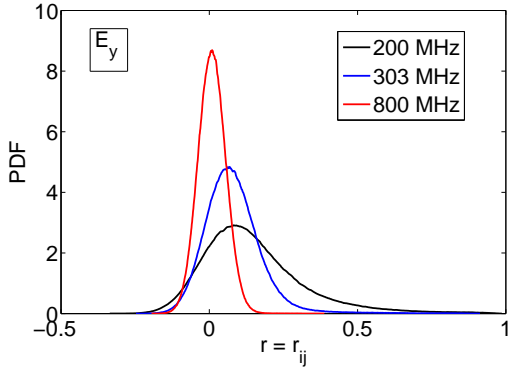}
\caption{Probability density function of the off-diagonal element of correlation matrix for $\left | E_y \right |$ at low (200~MHz),
intermediate (303~MHz), and high (800~MHz) frequency, for a spatial grid $13 \times 13 \times 11$ of the US--PM strategy.}
\label{fig:Ey_rij}
\end{figure}

\begin{figure}[!t]
\centering
\includegraphics[width=3in]{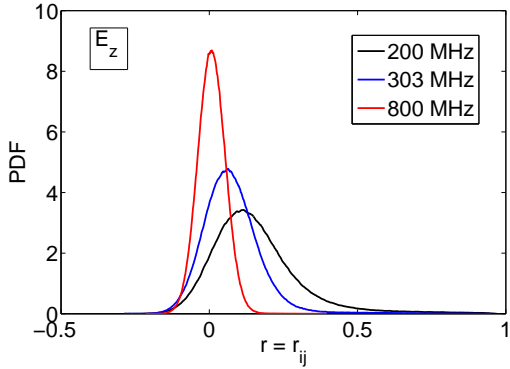}
\caption{Probability density function of the off-diagonal element of correlation matrix for $\left | E_z \right |$ at low (200~MHz),
intermediate (303~MHz), and high (800~MHz) frequency, for a spatial grid $13 \times 13 \times 11$ of the US--PM strategy.}
\label{fig:Ez_rij}
\end{figure}

The PDF of $f_{R} \left ( r \right )$ is reported in \figurename~\ref{fig:Ex_rij}
for the correlation of the Cartesian field $\left | E_x \right |$, 
as obtained through the US--PM strategy. 
Similar distributions are obtained for the components $E_y$ and $E_z$. Those are shown in \figurename~\ref{fig:Ey_rij}
and \figurename~\ref{fig:Ez_rij}. 
Interestingly, the PDF of $\left | E_x \right |$, $\left | E_y \right |$, and $\left | E_z \right |$ has non-central $t$-student profile. 
This is consistent with the statistics of two superimposed partially developed speckles in spatial optics \cite{2009_Yaitskova_speckle}.   
Also, observations confirm isotropic behavior and ergodicity over the investigated frequency range. 
Furthermore, it looks had to get rid of any residual correlation, whence the importance of fixing a threshold, and the 
genesis of the uncorrelation seems to be related to the positive tail. This has the important implication of changing the conventional 
uncorrelation threshold from $\left | r_{ij} \right | < \rho_e$ to $r_{ij} < \rho_e$. 
In this perspective, the presence of a threshold is of key importance as it looks that non-central behavior
of the correlation PDF at low frequencies 
creates a fat tail in the region $r_{ij} > e^{-1}$. 

\begin{figure}[!t]
\centering
\includegraphics[width=3in]{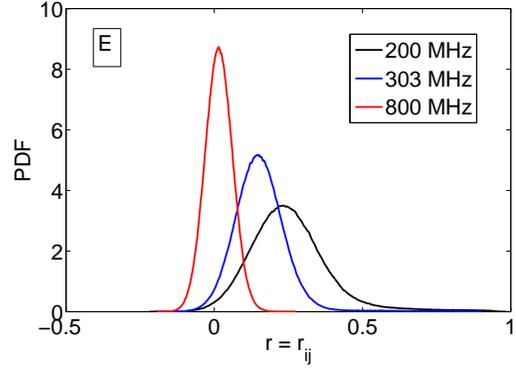}
\caption{Probability density function of the off-diagonal element of correlation matrix for $\left | E \right |$ at low (200~MHz),
intermediate (303~MHz), and high (800~MHz) frequency, for a spatial grid $13 \times 13 \times 11$ of the US--PM strategy.}
\label{fig:E_rij}
\end{figure}

In order to speculate more on the importance of a threshold, we present the PDF of the two-position correlation function 
calculated through the total field $E$. This is reported in \figurename~\ref{fig:E_rij}. 
Interestingly, we observe that the average value of the non-central $t$-student at low frequency is located 
where the tail of the high frequency distribution is almost absent. 
Furthermore, it looks that this average value goes to $e^{-1}$ as the frequency approaches the LUF, $f_l$.
This suggest the definition of a threshold which is strictly related to the RC under consideration.
In particular, the arbitrary choice of a threshold can be turned into  
a more ``physics-based'', or RC based, definition through 
\begin{equation}
 \overline{r} \approx \lim_{f \rightarrow f_l} \, \left < r_{\tau_i , \tau_j} \left ( f \right ) \right >_{\tau = \tau_i - \tau_j} \,\, , 
\end{equation} 
with $r_{\tau}$ calculated as the total field correlation over a proper sampling lattice. 

\subsection{Uncorrelated spatial points}
\label{sec:rescaso_up}
\begin{figure}[!t]
\centering
\includegraphics[width=2.5in]{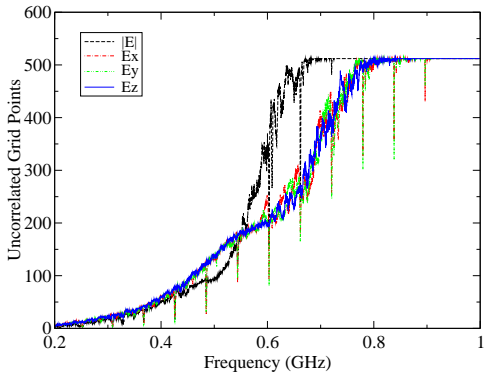}
\caption{Simulated uncorrelated grid points for a correlation matrix of stirrer positions (UP--SM).
         Each component and the magnitude of the electric field are reported.}
\label{fig:upsm}
\end{figure}

\begin{figure}[!t]
\centering
\includegraphics[width=2.5in]{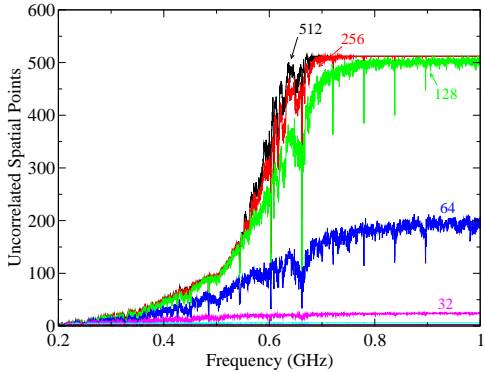}
\caption{Simulated uncorrelated grid points as function of the number of stirrer positions
         used in the computation of the correlation matrix (UP--SM). The $|E|$ was used.}
\label{fig:upsm_stirrer}
\end{figure}

\figurename~\ref{fig:upsm} shows the simulated the number of uncorrelated grid points
for a correlation matrix of stirrer angles (UP--SM).
Again, when computed through the total electric field magnitude,
this number exhibits higher values with respect to the single components.
By using the rectangular components, the adopted grid points are uncorrelated above 800~MHz, where the point distance
is $0.4 \lambda$. 
This result is close to the expected spatial correlation length of $0.5 \lambda$. 
By using the total field, the adopted grid points are uncorrelated above 650~MHz, where the point distance
is $0.325 \lambda$.
This could be explained by a consideration similar to that done and the end of Section \ref{sec:rescaso_uf}.
The effect of the number of stirrer positions used to compute the correlation
matrix is reported in \figurename~\ref{fig:upsm_stirrer}.
When the number of positions is large enough, the results converge to the same values,
while decreasing the number of positions gives rise to lower values.
This means that the correlation must be computed over a sufficiently large data ensemble.
Moreover, regular frequency peaks where correlation increases appear.
These peaks are also visible in the US--PM case and investigated in \cite{2013_EMCS_Correlazione}.
They do not depend on the adopted strategy, but on the RC geometry.
In particular, peak distance corresponds to a value for which the RC dimension along the stirrer rotation axis
is $0.5 \lambda$, which is 60~MHz for this RC.

\begin{figure}[!t]
\centering
\includegraphics[width=2.5in]{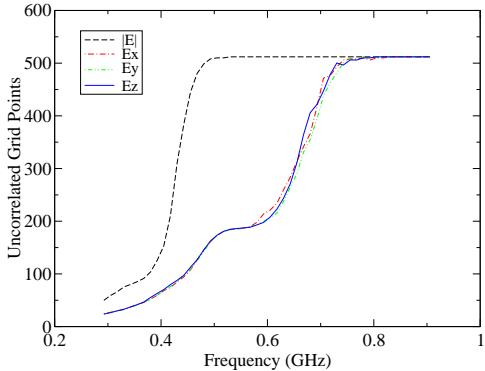}
\caption{Simulated uncorrelated grid points for a correlation matrix of frequency points (UP--FM).
         Each component and the magnitude of the electric field are reported.}
\label{fig:upfm}
\end{figure}

\begin{figure}[!t]
\centering
\includegraphics[width=2.5in]{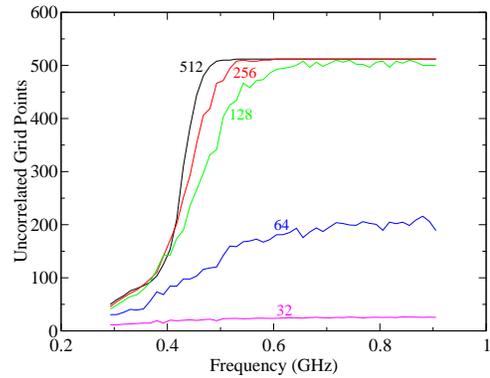}
\caption{Simulated uncorrelated grid points as function of the number of frequency points
         used in the computation of the correlation matrix (UP--FM). The $|E|$ was used.}
\label{fig:upfm_frequenze}
\end{figure}

\figurename~\ref{fig:upfm} shows the simulated number of uncorrelated grid points
for the strategy UP--FM.
At each working frequency, we compute the correlation coefficients using a set of $N_f$ frequency points,
which defines the frequency stirring bandwidth $B_w$.
When we use $N_f=512$, $B_w=156$~MHz. 
The effect of the number of stirrer positions on the estimated number of
uncorrelated frequencies is reported in \figurename~\ref{fig:upfm_frequenze}.
The frequency points where the correlation is calculated are the same.
All the frequency points are equidistant, therefore reducing $N_f$ has the effect of decreasing $B_w$.
When the number of frequency points is large enough, the results converge. 
Decreasing $N_f$, a reduction of the number of uncorrelated points is observed.
In the FM strategies, peaks where uncorrelation decreases are not visible due to the intrinsic frequency averaging of the method.

\section{Conclusions}
We have compared existing methods to estimate the number of uncorrelated stirrer positions
through the correlation matrix of multivariate stir traces. 
An actual reverberation chamber has been simulated through a full wave, finite-difference time-domain parallel code.
Spatial and frequency multi-point sampling of stir traces offer an overall picture
of the field distributions of the reverberation chamber. 
A threshold is introduced to count the uncorrelated pairs of stirrer positions.
Pertaining this case, we observed that the classical autocorrelation method overestimates the uncorrelated blade positions.
Estimation of uncorrelated frequencies and spatial grid points is also possible,
which is of interest in frequency and spatial stirring techniques. 
We noted an overestimation of uncorrelated pairs if frequency samples are used.
The use of Cartesian field components underestimate the uncorrelated frequency and space points.
Conversely, uncorrelated stirrer positions are not effected by this choice.
In general, a sufficiently large date ensemble is necessary to correctly compute the correlation.
Finally, the probability density function of the correlation matrix elements has been investigated. 
The observed frequency evolution (function width and positive tail) suggests a threshold value tailored
to the lowest usable frequency of the chamber.

\section*{Acknowledgment}
We acknowledge PRACE for awarding us access to resource FERMI based in Italy at CINECA.

\bibliographystyle{IEEEtran}

\end{document}